\documentclass[aps,prl,twocolumn,superscriptaddress]{revtex4-1}
\newcommand{\revision}[1]{{{#1}}}

\usepackage{graphicx}
\usepackage{dcolumn}
\usepackage{bm}
\usepackage{accents}
\usepackage{amssymb}
\usepackage{feynmp}
\usepackage{url}
\usepackage{setspace}
\usepackage{color}
\usepackage{hyperref}

\newcommand{\beq}{\begin{equation}}
\newcommand{\eeq}{\end{equation}}
\newcommand{\beqa}{\begin{eqnarray}}
\newcommand{\eeqa}{\end{eqnarray}}
\newcommand{\Tr}{\text{Tr}}

\begin{document}

\title{\revision{General bound on the performance of counter-diabatic driving acting on dissipative spin systems}}

\author{Ken Funo}
\affiliation{Theoretical Physics Laboratory, RIKEN Cluster for Pioneering Research, Wako-shi, Saitama 351-0198, Japan}
\author{Neill Lambert}
\affiliation{Theoretical Physics Laboratory, RIKEN Cluster for Pioneering Research, Wako-shi, Saitama 351-0198, Japan}
\author{Franco Nori}
\affiliation{Theoretical Physics Laboratory, RIKEN Cluster for Pioneering Research, Wako-shi, Saitama 351-0198, Japan}
\affiliation{RIKEN Center for Quantum Computing (RQC), Wako-shi, Saitama 351-0198, Japan}
\affiliation{Physics Department, The University of Michigan, Ann Arbor, Michigan 48109-1040, USA}

\date{\today}

\begin{abstract}
Counter-diabatic driving (CD) is a technique in quantum control theory designed to counteract nonadiabatic excitations and guide the system to follow its instantaneous energy eigenstates, and hence has applications in state preparation, quantum annealing, and quantum thermodynamics. However, in many practical situations, the effect of the environment cannot be neglected, and the performance of the CD is expected to degrade. To arrive at \revision{general} bounds on the resulting error of CD in this situation we consider a driven spin-boson model as a prototypical setup. The inequalities we obtain, in terms of either the Bures angle or the fidelity, allow us to estimate the maximum error solely characterized by the parameters of the system and the bath. By utilizing the analytical form of the upper bound, we demonstrate that the error can be systematically reduced through optimization of the external driving protocol of the system.  We also show that if we allow a time-dependent system-bath coupling angle, the obtained bound can be saturated and realizes unit fidelity. 
\end{abstract}

\maketitle

{\it Introduction.---} Counter-diabatic driving (CD) is a method to guide the system along a given adiabatic trajectory~\cite{STAreview,STAR,DR03,DR05,Berry09,Jarzynski13,Adolfo10,Deffner14,STAnonH} and to reproduce the target state expected from quantum adiabatic protocols in finite time, hence realizing Shortcuts To Adiabaticity (STA)~\cite{STAreview,STAR}. 
With the correct CD, one can speedup a desired quantum operation with unit fidelity, a result which is extremely useful in many applications that require fast high-performance quantum operations, such as 
quantum gate operations~\cite{Santos15,Santos16,YHChen}, quantum annealing~\cite{Adolfo12,Takahashi17,Hatomura18}, state preparation~\cite{expCD1,expCD2,expCD3,expCD5,Nori21}, transport~\cite{expCD4,expCD6} interferometory~\cite{interfero}, geometric pumping~\cite{Funo20,Takahashi20}, and heat engines~\cite{Adolfo14,Berakdar16,Lutz18,Deng18,Adolfo18,Funo19}.  

The rapid theoretical progress and promise of STA in such applications has motivated experimental implementations~\cite{expCD1,expCD2,expCD3,expCD4,expCD5,expCD6}, most of which are designed to control the system quickly enough such that the effect of the environment is suppressed. However, environmental effects cannot be completely neglected, and the performance of the CD technique, which was originally designed for isolated systems, is expected to degrade in realistic conditions. Motivated by this, several studies focused on the robustness of the CD under decoherence and noise~\cite{Sun16,Santos17,Levy18}, whereas others attempt to generalize the STA and CD to open systems~\cite{Villazon19,Alipour20,Dann19,Dupays20,Pancotti20}, hoping to find optimal drives in the presence of noise. However, a systematic way of understanding the controllability of open quantum systems has not been established yet, since limitations arise from the inevitable approximations in the analytical methods or numerical calculations used to study these complex situations. An exact analytical approach is needed to clarify the controllability set by the CD acting on the system only, and gain physical intuition about how we can decrease the error due to the environment as much as possible. 

\begin{figure}[t]
\begin{center}
\includegraphics[width=.45\textwidth]{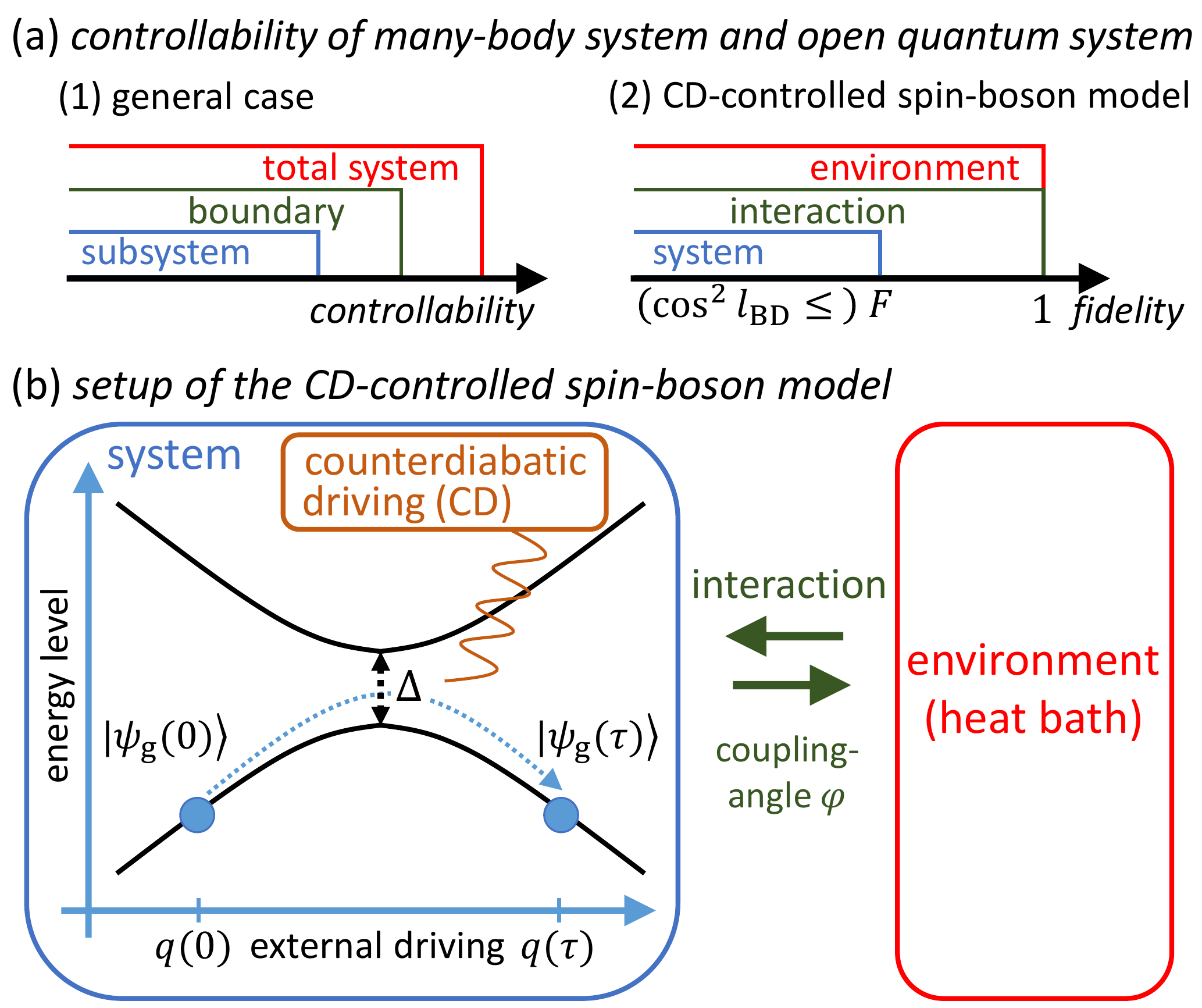}
\caption{\revision{Schematic diagrams of (a) the concept of levels of controllability and (b) the setup. (a.1) Controllability of a general many-body system. (a.2) Controllability of the spin-boson model. Unit fidelity is achieved by the exact STA protocol~(\ref{STA}) which requires a time-dependent control of the interaction. The lower bound on the fidelity $F$ set by the CD on the system alone is characterized by $\cos^{2} l_{\rm BD}$ via inequality~(\ref{Fidbound}). (b) We consider the spin-boson model to study the controllable limit of the spin system via CD under the influence of the environment. Here, controllability is measured by the ground state fidelity of the system. }
}
\label{fig_setup}
\end{center}
\end{figure}

It is expected that if one wishes to fully control the state of the system, engineering the system-environment coupling or the properties of the environment itself will become necessary [\revision{see Fig.~\ref{fig_setup}.~(a)}]. This opens up a connection to another interesting topic, the controllability of many-body systems, with possible applications to quantum adiabatic computing. It is known that constructing the CD 
requires precise knowledge about all instantaneous energy eigenstates. Even if this is possible, the resulting CD typically requires non-local interactions. 
To circumvent these points, recent studies aim to obtain approximate CD protocols~\cite{Campbell15,Polkovnikov17,Claeys19,Hatomura20}, akin to that needed for the open quantum systems we study in this work, and a method to estimate the error of the control would also be important in those approaches. 


In this Letter, we develop a \revision{general} bound on the performance of the CD under the influence of a heat bath by considering a driven spin-boson model [\revision{see Fig.~\ref{fig_setup}.~(b)}]. The spin-boson model is a prototypical minimal model describing a two-level system interacting with a continuum of bosonic bath modes, and is relevant for describing quantum information processing devices in a range of parameter regimes, from weak memory-less noise~\cite{Breuer,Lidar} to the non-Markovian, strong-coupling and non-rotating wave approximation regimes \cite{cqed1,cqed2,cqed3,cqed4,cqed5}. It is worth noting that even in the simplest case of the single-mode spin-boson (Rabi) model, integrability was a long standing issue and the exact solution was obtained only a decade ago~\cite{Rabi}. Therefore, one typically has to rely on numerical calculations, and apart from the seminal works in \cite{LZtrans1,LZtrans2}, little is known about exact analytical results for nonequilibrium dynamics in arbitrary parameter regimes. 
However, here we overcome this difficulty (for solving the spin-boson model) by utilizing powerful analytical tools such as the parallel transport via CD~\cite{Jarzynski13} and quantum speed limits (QSL)~\cite{MT,DC,Suzuki20}. We first show that by allowing a time-dependent system-bath coupling angle, we can construct {\it a unit fidelity protocol for obtaining the ground state of the system, realizing an exact STA}. We find that it is not necessary to have a full control of the environment in order to achieve the desired unit fidelity [see Fig.~\ref{fig_setup}.~(b.2)]. 
We next consider a more experimentally relevant situation where the system-bath coupling is static, and obtain {\it a lower bound on the fidelity when the system alone is controlled by the CD, which is the main result of our work}. Our result is \revision{general} in the sense that 
the result holds for arbitrary system Hamiltonian and bath spectral density.

{\it Counter-diabatic driving.---} To begin with, we consider an isolated Landau-Zener (LZ) 
model with CD. 
The total Hamiltonian is given by $H_{\rm cd}(t)=H_{0}(t)+H_{1}(t)$, where
\beq
H_{0}(t)=\frac{q(t)}{2}\sigma_{z}+\frac{\Delta}{2}\sigma_{x},  \ \ H_{1}(t)=\dot{\theta}_{t}\sigma_{y},
\eeq
and 
$\theta_{t}=(1/2)\cot^{-1}(q/\Delta)$, $\dot{\theta_{t}}=-\dot{q}\Delta/[2(\Delta^{2}+q^{2})]$.
Here, $H_{0}$ is the LZ Hamiltonian~\cite{LZreview}, where $\Delta$ characterizes the minimum gap, $q(t)$ describes the external driving, and $\sigma_{i}$ is the $i$-th component of the Pauli matrix. The CD Hamiltonian $H_{1}$ cancels non-adiabatic excitations and controls the system to stay in the instantaneous ground state $|\psi_{\rm g}(t)\rangle=\cos\theta_{t}\left|\downarrow\right\rangle-\sin\theta_{t}\left|\uparrow\right\rangle$ of $H_{0}$ during the unitary time-evolution generated by $H_{\rm cd}$. 

The mechanism of CD can be elegantly understood by the parallel transport argument~\cite{Jarzynski13}, but for later convenience, we explain the CD in terms of the unitary rotation $R_{t}=\exp(-i\theta_{t}\sigma_{y})$. 
After the unitary rotation, the CD Hamiltonian reads $R^{\dagger}_{t}H_{\rm cd}(t)R_{t}-iR^{\dagger}_{t}\partial_{t}R_{t}=\frac{1}{2}\sqrt{\Delta^{2}+q^{2}}\sigma_{z}$. Therefore, it is obvious that the system stays in the ground state $\left|\downarrow\right\rangle$ during the time-evolution in the rotated frame, which corresponds to $\left|\psi_{\rm g}(t)\right\rangle=R_{t}\left|\downarrow\right\rangle$ in the original frame, allowing the CD to parallel transport the system along $|\psi_{\rm g}(t)\rangle$. 

{\it Influence of the environment.---} We now analyze the influence of the environment on the CD by considering the spin-boson model, given by
\beq
H_{\varphi}(t)=H_{\rm cd}(t)+H^{\varphi}_{\rm int}+H_{B}. \label{sCD}
\eeq
Here, $H_{B}=\sum_{j}\omega_{k}b_{k}^{\dagger}b_{k}$ is the bath Hamiltonian describing a collection of harmonic oscillators, and $\omega_{k}$ and $b_{k}$ are the frequency and the annihilation operator of the $k$-th mode of the bath. The system-bath interaction Hamiltonian reads $H^{\varphi}_{\rm{int}}=\left(\cos2\varphi\sigma_{z}+\sin2\varphi\sigma_{x} \right) \otimes B$, 
where $\varphi$ is a coupling angle that determines in which direction the system-bath interaction mainly acts on, and 
$B=\sum_{k}g_{k}x_{k}$ shows that the system is linearly coupled to the ``position'' quadrature of the bath, where $g_{k}$ and $x_{k}=(b_{k}+b_{k}^{\dagger})/\sqrt{2\omega_{k}}$ are the coupling strength and the position operator of the $k$-th mode of the bath. The influence of the bath is fully characterized by the spectral density $J(\omega)=\pi \sum_{k}(g_{k}^{2}/2\omega_{k})\delta(\omega-\omega_{k})$, and we emphasize that our main result holds for arbitrary $J(\omega)$. 

We assume that the initial state of the composite system is given by the product state of the system ground state and the bath Gibbs state at inverse temperature $\beta$, i.e., $\rho(0) = |\psi_{\rm g}(0)\rangle\langle \psi_{\rm g}(0)| \otimes \rho_{B}^{\beta}$. Denoting the unitary time-evolution operator by $U_{\tau}=T\exp\left[ -i\int^{\tau}_{0}dt H_{\varphi}(t)\right] $, the final state of the composite system is given by $\rho(\tau)=U_{\tau}\rho(0)U_{\tau}^{\dagger}$. Since we are interested in the performance of the CD under the influence of the bath, we consider the fidelity $F=\langle \psi_{\rm g}(\tau)|\rho_{S}(\tau)|\psi_{\rm g}(\tau)\rangle$ between the target ground state $|\psi_{\rm g}(\tau)\rangle$ and the time-evolved state $\rho_{S}(\tau)=\Tr_{B}[\rho(\tau)]$. 
Note that for an isolated system, the CD is designed to obtain unit fidelity $F=1$, but this is no longer true when the system is influenced by the heat bath. 

{\it Exact STA via time-dependent coupling angle.---} Before deriving a bound on the fidelity, we discuss an interesting observation by using the unitary rotation $R_{t}$ that we introduced to explain the CD. Suppose that we allow a time-dependent rotation of the coupling angle $\varphi=\theta_{t}$ of the Hamiltonian~(\ref{sCD}), and denote it as $H_{\theta_{t}}(t)$. Then, in the rotated frame, we have $R^{\dagger}_{t}H_{\theta_{t}}(t)R_{t}-iR^{\dagger}_{t}\partial_{t}R_{t}=\frac{1}{2}\sqrt{\Delta^{2}+q^{2}}\sigma_{z}+\sigma_{z}\otimes B +H_{B}$. Note that this Hamiltonian in the rotated frame simply describes a pure dephasing effect from the bath, and the ground state of the system is unaffected during the time-evolution. In fact, we can obtain an explicit form of the time-evolved density matrix in the original frame, which is parallel transported along the ground state: 
\beq
\rho_{\rm sta}(\tau)=U_{\rm sta}\rho(0)U_{\rm sta}^{\dagger}=|\psi_{\rm g}(\tau)\rangle\langle \psi_{\rm g}(\tau)|\otimes \rho_{B}^{-}(\tau). \label{STA}
\eeq
Here, $U_{\rm sta}=T\exp[-i\int^{\tau}_{0}dt H_{\theta_{t}}(t)]$ is the time-evolution operator with $\varphi=\theta_{t}$, and $\rho_{B}^{-}(\tau)=e^{-iH_{B}^{-}\tau}\rho_{B}^{\beta}e^{iH_{B}^{-}\tau}$ is the time-evolved bath density matrix with respect to the 
position-shifted bath Hamiltonian $H_{B}^{-}=H_{B}-B$. 
In summary, the Hamiltonian~(\ref{sCD}) with the choice of $\varphi=\theta_{t}$ realizes an exact STA under the influence of the heat bath, i.e., the Hamiltonian transports the state of the system along its instantaneous ground state~(\ref{STA}) and realizes unit fidelity $F=1$.

It is interesting to note that controlling all of the bath degrees of freedom is unnecessary to achieve unit fidelity. Only a precise control of the coupling angle $\varphi=\theta_{t}$ is needed. Theoretically, this observation is important and has several advantages since the protocol and the time-evolved state have simple analytical expressions. In particular, we make use of this explicit form of the exact STA protocol to derive bounds on the performance of the CD on the system alone, i.e., with uncontrolled coupling angle $\varphi$, which is more relevant for most experimental situations where the coupling cannot be controlled directly. 

{\it \revision{General} bounds on the dissipative Landau-Zener CD.---} We now derive a lower bound on the fidelity of obtaining the target ground state for the CD under the influence of the heat bath. We first map the fidelity into the Bures angle defined as $\mathcal{L}(\rho,\sigma)=\arccos \sqrt{F(\rho,\sigma)}$~\cite{Nielsen,footnote2}. Here, the relation between $F$ and $\mathcal{L}$ is flipped: the Bures angle takes the minimal value $\mathcal{L}=0$ (the maximal value $\mathcal{L}=\pi/2$) when the fidelity takes the maximal value $F=1$ (the minimal value $F=0$). In what follows, we derive an upper bound on the Bures angle, which is later converted into a lower bound on the fidelity. 

We begin by using the contractivity of the Bures angle under partial trace of the bath degrees of freedom~\cite{Nielsen}. Then, the Bures angle between the target ground state and the CD-controlled state of the system can be bounded from above as
\beq
\mathcal{L}[|\psi_{\rm g}(\tau)\rangle,\rho_{S}(\tau)] \leq \mathcal{L}[\rho_{\rm sta}(\tau),\rho(\tau)], \label{CPTP}
\eeq
where $\rho_{\rm sta}$ is given in Eq.~(\ref{STA}). 
We then apply the quantum speed limit (QSL) inequality obtained by Suzuki and Takahashi~\cite{Suzuki20}, which in our case reads
\beq
\mathcal{L}[\rho_{\rm sta}(\tau),\rho(\tau)] \leq \int^{\tau}_{0}dt \sqrt{V_{\rho_{\rm sta}}[H_{\theta_{t}}-H_{\varphi}]}, \label{QSL1}
\eeq
where $V_{\sigma}[X]=\Tr[\sigma X^{2}]-(\Tr[\sigma X])^{2}$ is the variance. By following Ref.~\cite{Suzuki20}, the inequality~(\ref{QSL1}) is obtained from the standard QSL~\cite{MT,DC} as follows. The QSL gives an upper bound on the Bures angle between the initial and the final state in terms of the energy fluctuation: $\mathcal{L}[\sigma(\tau),\sigma(0)]\leq \int^{\tau}_{0}dt\sqrt{V_{\sigma}[H]}$. Here, the time-evolution of $\sigma(t)$ is generated by $H(t)$. Now, let us define $\tilde{X}(t)=U_{t}^{\dagger}X(t)U_{t}$. Then, the unitary invariance of the Bures angle reads $\mathcal{L}[\rho_{\rm sta}(\tau),\rho(\tau)]=\mathcal{L}[\tilde{\rho}_{\rm sta}(\tau),\tilde{\rho}_{\rm sta}(0)]$, by noting that $\tilde{\rho}_{\rm sta}(0)=\rho(0)$. Moreover, the time-evolution equation of $\tilde{\rho}_{\rm sta}$ reads $\partial_{t}\tilde{\rho}_{\rm sta}=-i[\tilde{H}_{\theta_{t}}-\tilde{H}_{\varphi},\tilde{\rho}_{\rm sta}]$. Therefore, by substituting $\sigma=\tilde{\rho}_{\rm sta}$ and $H=\tilde{H}_{\theta_{t}}-\tilde{H}_{\varphi}$ inside the QSL and noting $V_{\tilde{\sigma}}[\tilde{X}]=V_{\sigma}[X]$, we obtain~(\ref{QSL1}). 


Note that Ref.~\cite{Suzuki20} applied the QSL to obtain a bound on the performance of adiabatic quantum computation, whereas we are here interested in quantifying the performance of the CD under the influence of the bath.

Now, the explicit and simple form of $\rho_{\rm sta}$ given in Eq.~(\ref{STA}) allows us to analytically calculate the right-hand side of~(\ref{QSL1}). First of all, $H_{\theta_{t}}-H_{\varphi}=\Delta H^{S}_{\rm int}\otimes B$ with $\Delta H^{S}_{\rm int}=(\cos2\theta_{t}-\cos2\varphi)\sigma_{z}+(\sin2\theta_{t}-\sin2\varphi)\sigma_{x}$. Therefore, the variance in Eq.~(\ref{QSL1}) reads 
\beqa
V_{\rho_{\rm sta}} [H_{\theta_{t}}-H_{\varphi}] &=& \langle \psi_{\rm g}| (\Delta H^{S}_{\rm int})^{2}|\psi_{\rm g}\rangle \Tr[B^{2}\rho_{B}^{-}(t)] \nonumber \\
& &- \langle \psi_{\rm g}|\Delta H^{S}_{\rm int}|\psi_{\rm g}\rangle^{2} (\Tr[B\rho^{-}_{B}(t)])^{2}. \label{Varsta}
\eeqa
The system-dependent part in~(\ref{Varsta}) can be easily obtained as $ \langle \psi_{\rm g}| (\Delta H^{S}_{\rm int})^{2}|\psi_{\rm g}\rangle =4\sin^{2}(\theta_{t}-\varphi)=-2\langle \psi_{\rm g}|\Delta H^{S}_{\rm int}|\psi_{\rm g}\rangle$. 
The bath-dependent part reads~\cite{footnote1} $\Tr[B\rho_{B}^{-}(t)]=X_{t}$, where $X_{t}=\int d\omega (2/\pi\omega) J(\omega)(1-\cos\omega t)$ quantifies the expectation value of the (coupling-constant multiplied) bath position that is shifted by $H_{B}^{-}$. Also, $\Tr[B^{2}\rho_{B}^{-}(\tau)]=S+X_{t}^{2}$, where $S=\int d\omega (J(\omega)/\pi) \coth(\beta\omega/2)$ is the bath correlation function $\langle B(t)B(0)\rangle$ at $t=0$. We further note that $S$ scales as $O(\beta^{-1})$ in the high-temperature limit, whereas it is the integrated spectral density in the zero-temperature limit. 

We now obtain our main result by combining~(\ref{CPTP}), (\ref{QSL1}) and (\ref{Varsta}), which gives an upper bound on the Bures angle between the target ground state and the CD-controlled state of the system:
\beq
\mathcal{L}[|\psi_{\rm g}(\tau)\rangle,\rho_{S}(\tau)] \leq l_{\rm BD} \label{varphibound}
\eeq
with
\beq
l_{\rm BD}= \int^{\tau}_{0}dt |2\sin(\theta_{t}-\varphi)|\sqrt{ S + \cos^{2}(\theta_{t}-\varphi)X^{2}_{t}}. \label{boundl}
\eeq
Here, the left-hand side of~(\ref{varphibound}) quantifies the error of the CD, since a small value of $\mathcal{L}$ means that the CD-controlled state $\rho_{\rm S}(\tau)$ is close to the target ground state. The bound $l_{\rm BD}$ gives a \revision{general} upper bound on the error, in the sense that it does not require information about the actual nonequilibrium dynamics of the system. As we see from Eq.~(\ref{boundl}), $l_{\rm BD}$ depends only on predefined quantities, such as the driving protocol $q(t)$ through $\theta_{t}$, the coupling angle $\varphi$, and the bath properties $S$ and $X_{t}$. The error becomes larger as either the system-bath coupling strength (i.e., $S$ and $X_{t}$) becomes larger or the driving protocol is unoptimized, such that $\theta_{t}$ deviates from $\varphi$.

Note that the bound~(\ref{varphibound}) is tight and can be saturated by the exact STA protocol ($\varphi=\theta_{t}$) with Eq.~(\ref{STA}). For the general case, the analytical form of the upper bound allows us to optimize the parameters through minimizing $l_{\rm BD}$, and increase the performance of the CD. Later, in the {\it applications}, we demonstrate the usefulness of the bound~(\ref{varphibound}) by optimizing the driving protocol $q(t)$.  

Since the maximum value of the Bures angle is given by $\pi/2$, the bound~(\ref{varphibound}) is meaningful when $l_{\rm BD} \leq \pi/2$. In such cases, we can convert the inequality~(\ref{varphibound}) into a lower bound on the fidelity, given by
\beq
F[|\psi_{\rm g}(\tau)\rangle,\rho_{S}(\tau)]  \geq \cos^{2} l_{\rm BD}. \label{Fidbound}
\eeq
To summarize, both inequalities~(\ref{varphibound}) and~(\ref{Fidbound}) quantify the performance of the CD under the influence of the heat bath. In the following, we give several additional comments on our results. First, it is straightforward to generalize the result to the case of obtaining the excited state, or a classical mixture of the ground and excited states, whereas we find that the upper bound~(\ref{boundl}) is unchanged and~(\ref{varphibound}) is still valid. 
Second, we discuss the dependence of the system-bath coupling strength $\lambda$ on the bound. 
Since $S=O(\lambda^{2})$ and $X_{t}^{2}=O(\lambda^{4})$, 
the inequality~(\ref{Fidbound}) becomes $F\geq 1-4S[\int^{\tau}_{0}dt|\sin(\theta_{t}-\varphi)|]^{2}+O(\lambda^{4})$ in the weak-coupling limit, and the discrepancy from unit fidelity scales quadratically with $\lambda$. 
In addition, by using reservoir engineering, one can in principle engineer the bath spectral density to reduce  $S$, suppressing the CD error.  

\begin{figure}[t]
\begin{center}
\includegraphics[width=.45\textwidth]{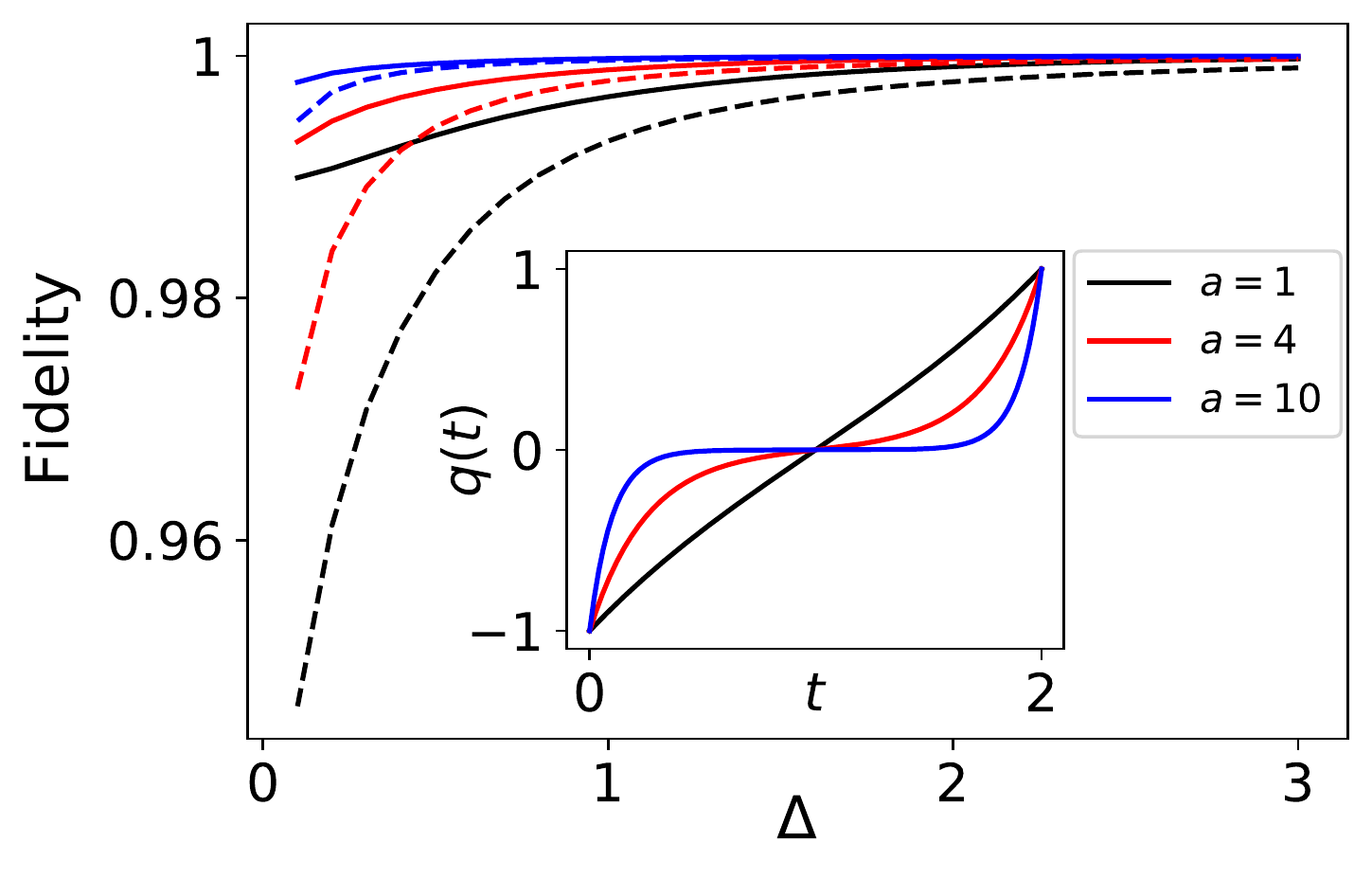}
\caption{ Numerical demonstration of the bound~(\ref{Fidbound}) by varying $\Delta$. The solid curves show the fidelity $F[|\psi_{\rm g}(\tau),\rho_{S}(\tau)]$ and the dashed curves are the lower bound $\cos^{2}l_{\rm BD}$~(\ref{boundl}) for different values of $a$. \revision{The inset shows the functional form of the external driving $q(t)= \sinh(a(t-\tau/2))/\sinh(a\tau/2)$, which is designed to better approximate the optimal drive $q(0)=-1$, $q(t)=0$  $(0<t<\tau)$, and $q(\tau)=1$, as $a$ increases. Therefore,} the lower bound $\cos^{2}l_{\rm BD}$ and also the fidelity becomes closer to unity as $a$ becomes larger, demonstrating the effectiveness of our inequality~(\ref{Fidbound}). The parameters are $\varphi=\pi/4$ ($\sigma_{x}$-coupling), $\beta=1,\gamma=0.1, w_{0}=1,\lambda=0.1, \tau=2$. 
}
\label{fig_bound}
\end{center}
\end{figure}

{\it Applications.---} 
\revision{We now consider finding a protocol $q(t)$ that would give better fidelity by reducing $l_{\rm BD}$~(\ref{boundl}). 
We assume that the initial and final values of $q(t)$ are fixed, i.e., $q(0)=q_{\rm i}$ and $q(\tau)=q_{\rm f}$, but at intermediate times, $q(t)$ is unfixed. Then, the optimal drive $q(t)$ that minimizes $l_{\rm BD}$ is given by $q(0)=q_{\rm i}$, $q(t)=q_{*}$ $(0<t<\tau)$, and $q(\tau)=q_{\rm f}$, where $q_{*}=\Delta\cot(2\varphi)$. To show this claim, we discretize the time-integral in Eq.~(\ref{boundl}) with $\Delta t$ being the time-duration of one step and $N$ being the total number of steps, i.e., $\tau=N\Delta t$, We denote $f[q(t)]$ as the integrand given in Eq.~(\ref{boundl}) and use the property $f[q_{*}]=0$ to obtain $l_{\rm BD}=f[q_{\rm i}]\Delta t+f[q_{\rm f}]\Delta t+O(\Delta t^{2})\rightarrow 0$ as $\Delta t\rightarrow 0$, and thus $q(t)$ given above is optimal. 

As a concrete example, we set $q_{\rm i}=-1$ and $q_{\rm f}=1$ and assume a $\sigma_{x}$ coupling ($\varphi=\pi/4$). Note that the optimal drive requires sudden changes of the drive at inital and final times, causing the CD control field $\dot{\theta}_{t}\propto \dot{q}(t)$ to diverge. To circumvent this point, we consider the smooth functional form $q(t)=\sinh(a(t-\tau/2))/\sinh(a\tau/2)$ to approximate the optimal drive. For larger $a$, this becomes a better approximation to the the optimal drive,} and the lower bound on the fidelity $\cos^{2}l_{\rm BD}$ becomes larger, as plotted by dashed curves in Fig.~\ref{fig_bound}. It is worth noting that the actual performance of the CD, measured by the fidelity, becomes also better for large $a$ (solid curves), suggesting the practical usefulness of the bound~(\ref{Fidbound}). Here, the numerical calculation is performed using the hierarchal equations of motion (HEOM) method~\cite{Tanimurareview} implemented in the BoFiN extension \cite{Lambert19,Bofin}  for QuTiP~\cite{Qutip1,Qutip2}, where the following under-damped Brownian motion spectral density is used: $J(\omega)=\gamma\lambda^{2}\omega/[(\omega^{2}-\omega_{0}^{2})^{2}+\gamma^{2}\omega^{2}]$.
Here, $\omega_{0}$, $\gamma$, and $\lambda$ are the resonance frequency, width, and system-bath coupling strength, respectively. 

{\it Generalizations.---} Finally, we discuss generalizations of~(\ref{varphibound}) to multiple heat baths $H_{B}=\sum_{i}H_{B}^{i}$, an arbitrary system Hamiltonian $H_{0}(t)$, and a system-bath interaction $H_{\rm int}=\sum_{i}A_{i}\otimes B_{i}$, where $A_{i}$ is an arbitrary operator acting on the system and $B_{i}$ is the operator $B$ defined previously for the $i$-th bath. The total Hamiltonian is given by $H(t)=H_{0}(t)+H_{1}(t)+H_{\rm int}+H_{B}$, where $H_{1}(t)$ is the CD Hamiltonian for $H_{0}(t)$. We take the $n$-th energy eigenstate $|\psi_{n}(0)\rangle$ of $H_{0}(0)$ as the initial state of the system. An exact STA can be constructed by choosing the time-dependent system-bath interaction $H^{\rm sta}_{\rm int}(t)=-|\psi_{n}(t)\rangle\langle \psi_{n}(t)|\otimes \sum_{i}B_{i}$. 
By following a derivation similar to that for~(\ref{varphibound}), we obtain an upper bound on the Bures angle as $\mathcal{L}[ |\psi_{n}(\tau)\rangle,\rho_{\rm S}(\tau)]\leq l_{\rm BD}=\int^{\tau}_{0}dt \sqrt{g}$, with
\beq
g=\sum_{i}\langle \psi_{n}|(A_{i}+I)^{2}|\psi_{n}\rangle  S_{i} + \sum_{i,j}\text{Cov}_{|\psi_{n}\rangle}(A_{i},A_{j})X^{i}_{t}X_{t}^{j} \label{generalBD}.
\eeq
Here $I$ is the identity matrix of the system, $\text{Cov}_{|\phi\rangle}(A,B):=\langle \phi|AB|\phi\rangle-\langle \phi|A|\phi\rangle\langle \phi|B|\phi\rangle$ is the covariance, and $S_{i}$ and $X_{t}^{i}$ are $S$ and $X_{t}$ defined previously for the $i$-th bath. Note that similar to~(\ref{boundl}), the generalized bound depends solely on the system properties $A_{i}$ and $|\psi_{n}(t)\rangle$ and the bath properties $S_{i}$ and $X^{i}_{t}$. In addition, we note that Eq.~(\ref{generalBD}) reproduces the bound~(\ref{boundl}) for the LZ model~(\ref{sCD}) with the target state being $|\psi_{\rm g}(t)\rangle$.

{\it Conclusions.---} We have derived \revision{general} bounds on the performance of the CD under the influence of an environment by considering the spin-boson model. 
The upper bound on the error of the CD does not depend on the time-evolved state of the system, and is solely characterized by the parameters of the system and the bath. 
The obtained bound is tight and can be saturated by allowing a time-dependent system-bath coupling angle, realizing unit fidelity, and we call this protocol as an exact STA protocol. 
Our work clarifies the controllable limit via CD, and has immediate impact on current quantum information processing experiments by providing tools for error estimation and parameter optimization.  
Generalizations of our main result to arbitrary system Hamiltonian have been discussed, and further extensions to characterize the controllable bound and control error in generic many-body systems via approximate CD protocols would be an interesting direction of research.


\begin{acknowledgments}
{\it Acknowledgements.---} We thank K. Saito for useful discussions and comments. K.F.~was supported by the JSPS KAKENHI Grant Number JP18J00454.  N.L.~acknowledges partial support from JST PRESTO through Grant No.~JPMJPR18GC. F.N. is supported in part by: NTT Research, Japan Science and Technology Agency (JST) (via the Q-LEAP program, Moonshot R\&D Grant No. JPMJMS2061, and the CREST Grant No. JPMJCR1676), Japan Society for the Promotion of Science (JSPS) (via the KAKENHI Grant No. JP20H00134 and the JSPS-RFBR Grant No. JPJSBP120194828), Army Research Office (ARO) (Grant No. W911NF-18-1-0358), Asian Office of Aerospace Research and Development (AOARD) (via Grant No. FA2386-20-1-4069). F.N.~, N.L.~and K.F.~ acknowledge the Foundational Questions Institute Fund (FQXi) via Grant No. FQXi-IAF19-06.
\end{acknowledgments}


\begin{thebibliography}{99}
\bibitem{footnote2} The fidelity between mixed states is defined as~\cite{Nielsen} $F(\rho,\sigma)=(\Tr[\sqrt{\rho^{1/2}\sigma \rho^{1/2}}])^{2}$.
\bibitem{footnote1} 
We note that $B$ in the interaction picture reads $B^{-}(t)=e^{iH_{B}^{-}t}Be^{-iH_{B}^{-}t}=B(t) +X_{t}$ with $B(t)=e^{iH_{B}t}Be^{-iH_{B}t}$. Therefore, we obtain $\langle B^{-}(t)\rangle=X_{t}$ and $\langle B^{-}(t)^{2}\rangle = \langle B(t)^{2}\rangle +X_{t}^{2}=\langle B^{2}\rangle +X_{t}^{2}=S+X_{t}^{2}$.
\end{thebibliography}
\end{document}